\documentclass[lettersize,journal]{IEEEtran}
\usepackage{amsmath,amsfonts}
\usepackage{algorithmic}
\usepackage{algorithm}
\usepackage{array}
\usepackage{caption}
\usepackage{graphics}
\usepackage{subcaption}

\usepackage{textcomp}
\usepackage{stfloats}
\usepackage{url}
\usepackage{verbatim}
\usepackage{graphicx}
\usepackage{xspace,xcolor,colortbl}
\usepackage{hyperref}
\usepackage{multirow}
\usepackage[final]{microtype}
\usepackage{enumitem}

\usepackage{cite}

\usepackage[para]{threeparttable}
\usepackage{etoolbox}
\AtEndEnvironment{threeparttable}{\vskip10pt}{}{}
\makeatletter
\patchcmd{\TPT@doparanotes}
  {\TPT@hsize}
  {\TPT@hsize\footnotesize}
  {}
  {}
\makeatother

\colorlet{myGray}{gray!40}

\newcommand{\design}{FSL-HDnn\xspace}

\newcommand{\hd}{HDC\xspace}

\def\eg{\emph{e.g.}\xspace}
\def\ie{\emph{i.e.}\xspace}

\usepackage{titlesec}
\titlespacing*{\section}{0pt}{1ex plus 0.6ex minus .2ex}{.4ex plus .1ex}
\titlespacing*{\subsection}{0pt}{1ex plus .1ex minus .0ex}{.1ex plus .0ex}
\titlespacing*{\subsubsection}{0pt}{1ex plus 1ex minus 1ex}{1ex plus 1ex}

\renewcommand{\thesubsubsection}{\arabic{subsubsection}}
\titleformat{\subsubsection}[runin]{\itshape}{\thesubsubsection)}{1em}{}
\titlespacing*{\subsubsection}{\parindent}{0pt}{*1}

\begin{document}

\title{\design: A 40~nm Few-shot On-Device Learning Accelerator with Integrated Feature Extraction and Hyperdimensional Computing\vspace{-0.15cm}}

\author{
    Weihong Xu$^*$,~\IEEEmembership{Member, IEEE}, Chang Eun Song$^*$,~\IEEEmembership{Graduate Student Member, IEEE}, \\Haichao Yang,
    Leo Liu, Meng-Fan Chang,~\IEEEmembership{Fellow,~IEEE}, Carlos H. Diaz,~\IEEEmembership{Fellow,~IEEE}, \\Tajana Rosing,~\IEEEmembership{Fellow,~IEEE}, and  Mingu Kang,~\IEEEmembership{Member,~IEEE}
    \vspace{-0.95cm}
    \thanks{$^*$The authors contribute equally.}
    \thanks{This work was supported by TSMC and in part by PRISM and CoCoSys, centers in JUMP 2.0, an SRC program sponsored by DARPA. }
    \thanks{Weihong Xu, Chang Eun Song, Haichao Yang, and Tajana Rosing are with the Department of Computer Science and Engineering, University of California San Diego, La Jolla, CA 92093 USA. (e-mail: \{wexu, cesong, tajana\}@ucsd.edu)}
    \thanks{Mingu Kang is with the Department of Electrical and Computer Engineering, University of California San Diego, San Diego, CA 92161 USA. (e-mail: m7kang@ucsd.edu)}
    \thanks{Leo Liu, Meng-Fan Chang, and Carlos H. Diaz are with Taiwan Semiconductor Manufacturing Company (TSMC).}
}



\maketitle

\begin{abstract}
    This paper introduces \design, an energy-efficient accelerator that implements the end-to-end pipeline of feature extraction and on-device few-shot learning (FSL). The accelerator addresses fundamental challenges of on-device learning (ODL) for resource-constrained edge applications through two synergistic modules: a parameter-efficient feature extractor employing weight clustering and an FSL classifier based on hyperdimensional computing (HDC). The feature extractor exploits weight clustering mechanism to reduce computational complexity, while the HDC-based FSL classifier eliminates gradient-based back propagation operations, enabling single-pass training with substantially reduced latency.
    Additionally, \design enables low-latency ODL and inference via two proposed optimization strategies, including an early-exit mechanism with branch feature extraction and batched single-pass training that improves hardware utilization. 
    Measurement results demonstrate that our chip fabricated in a 40~nm CMOS process delivers superior training energy efficiency of 6~mJ/image and end-to-end training throughput of 28~images/s on a 10-way 5-shot FSL task. The end-to-end training latency is also reduced by $2\times$ to $20.9\times$ compared to state-of-the-art ODL chips. 
   
\end{abstract}

\begin{IEEEkeywords}
    On-device learning, few-shot learning, hyperdimensional computing, energy-efficient accelerator, edge inference.
\end{IEEEkeywords}

\vspace{-0.5cm}
\section{Introduction}

\IEEEPARstart{E}{}\MakeLowercase{dge} intelligence is becoming increasingly pervasive, placing growing demands on  adaptation capabilities of deep neural network (DNN) models for user-specific personalization. On-device learning (ODL)~\cite{dhar2021odlsurvey} addresses this need by enabling local model training for dynamic environments and use cases, thereby bridging the gap between global model generalization and user-specific needs.
ODL offers a range of key benefits. First, it preserves user privacy  by keeping the data in the local device without offloading it to centralized data centers. Second, it enables low-latency learning, allowing edge devices to respond in real time with contextually relevant outputs. Third, it enhances energy efficiency by avoiding data transmission to the server, thereby eliminating the need for continuous network connectivity, which is a critical advantage in power-constrained scenarios such as IoT and edge environments.

\begin{figure}[t]
    \centering
    \includegraphics[width=1\linewidth]{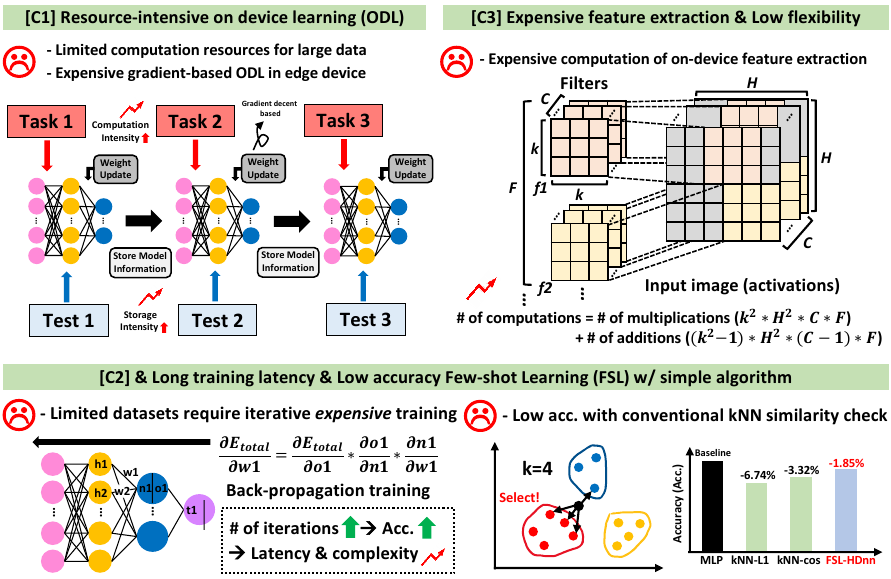}
    \caption{Challenges for existing on-device learning accelerators.}
    \label{fig:challenges}
\end{figure}

Despite these potential benefits, existing ODL accelerators face three critical challenges summarized as \textbf{C1}-\textbf{C3} in Fig.~\ref{fig:challenges}. State-of-the-art (SOTA) ODL accelerators\cite{han2021dflnpu,park2021neural,prabhu2022chimera,wang2022trainer,venkataramanaiah202328,qian20244,lee2021overview,heo2024sp,cai2020tinytl,heo2022t,tsai2023scicnn,liu2024high} generally adopt either full \cite{han2021dflnpu,park2021neural,wang2022trainer,venkataramanaiah202328,qian20244} or partial fine-tuning (FT)~\cite{prabhu2022chimera,heo2024sp,cai2020tinytl,liu2024high} to learn all or subsets of model weights with expensive gradient-based operations, respectively. But edge devices often operate under stringent \textbf{\underline{[C1]}  resource constraints for ODL}. Traditional DNN training approaches, relying on extensive back-propagation computations and large datasets, are impractical for resource-constrained environments because the computational overhead, memory consumption, and energy requirements associated with gradient-based training exceed the typical capabilities of edge devices. To this end, various software and hardware optimization techniques, including data sparsity exploitation~\cite{wang2022trainer,venkataramanaiah202328}, optimized dataflows~\cite{qian20244,han2021dflnpu,11173642},  emerging processing-in-memory solutions\cite{heo2022t,heo2024sp,song2025hybrid},  and non-volatile memory utilization~\cite{prabhu2022chimera}, are presented to reduce the training overhead. While these efforts alleviate the costly training overheads to some extent, they still suffer from the fundamental inefficiencies of gradient-based training, which demands high-precision arithmetic, sophisticated dataflows, and iterative operation with multiple epochs.

The distinctive challenge of edge ODL lies in the scarcity of training data due to limited user inputs~\cite{zhou2021device}. Unlike conventional DNN training that relies on large-scale datasets, edge scenarios typically have access to a limited number of user-specific training samples. This scenario introduces \textbf{\underline{[C2]} slow convergence and prolonged training latency} as gradient-based ODL methods have long convergence time and inadequate accuracy under limited data scenarios (see analysis in Section~\ref{sec:background}). Alternative approaches, such as k-nearest neighbor (kNN)~\cite{mao2022experimentally,li2021sapiens,karunaratne2022memory}, although computationally simpler without expensive back-propagation, result in suboptimal accuracy. Existing ODL accelerators also have \textbf{\underline{[C3]} expensive feature extractor with limited flexibility.} They predominantly utilize convolutional neural networks (CNNs) as the backbone feature extractors for the vision applications due to their effectiveness in handling complex tasks such as image classification~\cite{prabhu2022chimera} and object detection~\cite{du2018understanding}. The computation of CNN-based feature extraction constitutes a substantial portion of the overall computing workload, making it highly resource-intensive.




These challenges necessitate a significantly more efficient ODL framework to overcome existing limitations in latency, energy consumption, and bandwidth. Few-shot learning (FSL), a paradigm that enables models to adapt to new tasks using only a few labeled examples~\cite{ravi2017optimization}, offers a compelling solution by reducing reliance on large-scale datasets and costly retraining. To this end, we propose \design, an efficient ODL accelerator that integrates FSL with hyperdimensional computing (HDC). Inspired by brain-like computational principles~\cite{kanerva2009hyperdimensional,thomas2021theoretical}, HDC serves as a lightweight FSL classifier that supports extremely low-cost on-device learning while maintaining high accuracy. By keeping the feature extractor frozen and performing classification using HDC, our architecture is particularly suited for resource-constrained edge platforms where energy efficiency and adaptability are critical.
Our detailed contributions include:
\begin{enumerate}[leftmargin=*]
    \item We propose a novel single-pass and gradient-free FSL algorithm that eliminates expensive back propagation operations during training. By integrating HDC with transfer learning principles, the proposed \design algorithm significantly reduces on-device training complexity. 
    \item We present the \design architecture with two hardware-software co-optimizations: (1) a parameter-efficient feature extractor by exploiting weight clustering and a tailored dataflow; (2) a memory-efficient HDC-based FSL classification featuring cyclic random projection encoding.
    \item For low-latency edge deployment, we propose two optimization strategies: (1) an early exit mechanism with branch feature extraction that adaptively terminates inference at intermediate layers when feasible, and (2) batched single-pass training that processes multiple samples concurrently. 
    \item We fabricated the \design chip~\cite{yang2024fsl} in a 40~nm technology integrating the proposed optimizations.
    The measured results on a 10-way 5-shot few-shot ODL task demonstrate \design's superior energy efficiency (6 mJ/image with 2.9~TOPS/W), high throughput (28 images/s), and significant 2-20.9$\times$ end-to-end FSL speedup over existing ODL chips~\cite{han2021dflnpu, park2021neural, prabhu2022chimera, wang2022trainer, venkataramanaiah202328, qian20244} while maintaining competitive accuracy.  
\end{enumerate}

\begin{figure}[t]
    \centering
    \includegraphics[width=0.83\linewidth]{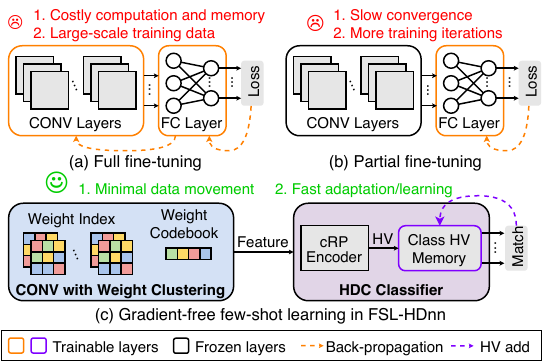}
    \caption{Comparison for different on-device learning algorithms (ODL): (a) full and (b) partial fine-tuning (FT), (c)  learning pipeline in the proposed gradient-free FSL-HDnn architecture.}
    \label{fig:odl_flow_comp}
\end{figure}

\section{Background}\label{sec:background}

\subsection{Existing ODL Algorithms}
Existing ODL accelerators primarily adopt gradient-based fine-tuning (FT) methods, categorized into full FT and partial FT as depicted in Fig.~\ref{fig:odl_flow_comp}(a) and (b). 

\noindent \textbf{1) Full fine-tuning}
in Fig.~\ref{fig:odl_flow_comp}(a) re-trains all model weights via four steps: forward propagation (FP), backward propagation (BP), gradient calculation (GC), and weight update (WU). For a classification task with $N_{\text{sample}}$ training samples and $T_{\text{itr}}$ iterations, the complexity is estimated as: 
\begin{equation}\label{eq:odl_full_cost}
    \small
    \text{Cost}_{\text{full}} \approx T_{\text{itr}} \times N_{\text{sample}} \times (\text{Cost}_{\text{FP}} + \text{Cost}_{\text{GC}} + \text{Cost}_{\text{BP}} + \text{Cost}_{\text{WU}}).
\end{equation}
While effective in cloud settings, the full FT approach incurs costly activation and training data storage, requires complex data flows such as weight transposition~\cite{heo2024sp}, and often suffers from poor generalization under limited data.

\begin{figure}[t] 
    \centering
    \includegraphics[width=0.95\linewidth]{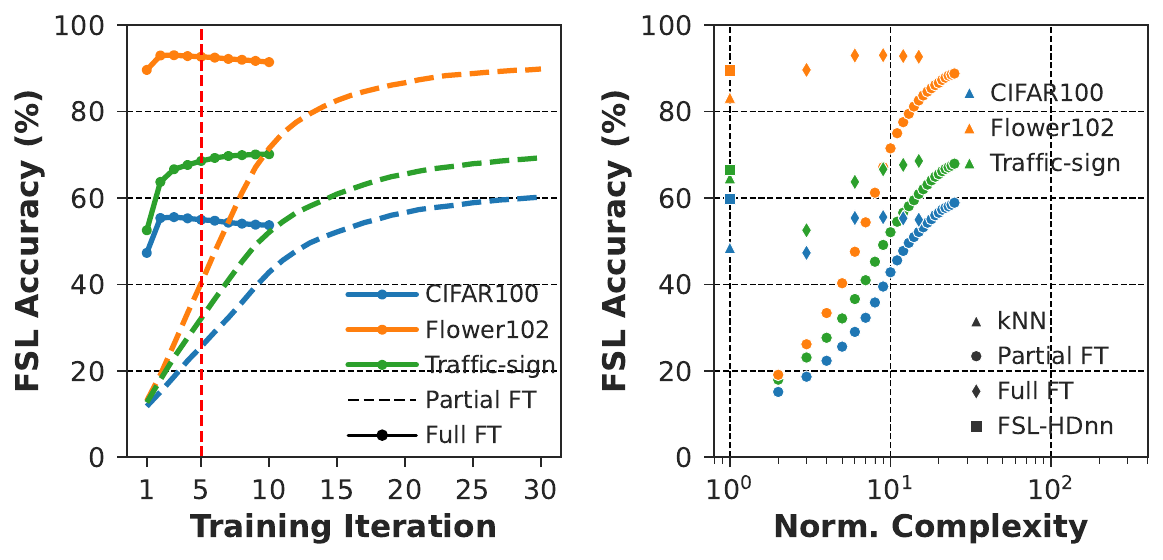}
    \caption{(a) FSL accuracy vs. training iterations for partial and full fine-tuning (FT) models and (b) accuracy vs. complexity (normalized to the smallest one) of kNN, partial FT, full FT, and \design algorithm.}
    \label{fig:convergence_complexity}
\end{figure}

\noindent \textbf{2) Partial fine-tuning} in Fig.~\ref{fig:odl_flow_comp}(b), achieves greater efficiency by reducing the high computational cost of full FT, training only a subset of weights while keeping the remaining weights 
frozen~\cite{cai2020tinytl,prabhu2022chimera,heo2024sp}. In this case, most BP and WU operations can be avoided~\cite{heo2024sp}, thereby reducing training cost to:
\begin{equation}\label{eq:odl_partial_cost}
    \text{Cost}_{\text{partial}} \approx T_{\text{itr}} \times N_{\text{sample}} \times (\text{Cost}_{\text{FP}} + \text{Cost}_{\text{GC}}).
\end{equation}
Despite the reduced cost and simplified dataflow by avoiding BP,  partial FT often requires much more iterations with a large $T_{itr}$ than full FT, resulting in prolonged latency and energy consumption for resource-constrained devices. 
This trend is well observed in Fig.~\ref{fig:convergence_complexity}, which depicts the training convergence and cost (in terms of normalized complexity) of popular ODL algorithms  under 20-way 5-shot FSL tasks\footnote{In this work, the $N$-way $k$-shot FSL task denotes an unseen $N$-class classification task to the model with $k$ training samples for each class.}.

To improve efficiency, kNN-based FSL methods\cite{li2021sapiens,mao2022experimentally} replace the final layer with a kNN classifier, thereby eliminating the need for gradient-based training.
As shown in Fig.~\ref{fig:convergence_complexity}(b), kNN-based approaches achieve orders of magnitude lower computational complexity compared to both full and partial FT.
However, kNN suffers from noticeable accuracy degradation relative to FT as shown in Fig.~\ref{fig:convergence_complexity}(b). These limitations motivate us to pursue a low-cost training approach with minimal iterations while preserving the accuracy of gradient-based methods.


\subsection{Hyperdimensional Computing (HDC)}
HDC~\cite{thomas2021theoretical,kanerva2009hyperdimensional} is a brain-inspired computational paradigm that operates on high-dimensional vectors, enabling efficient similarity-based learning and inference without requiring gradient computation or iterative training. 

\subsubsection{Encoding:} 
The key part of HDC is to encode a feature vector $\mathbf{x} \in \mathbb{R}^F$ into a $D$-dimensional hypervector (HV) $\mathbf{h} \in \mathbb{R}^D$ through the encoding process.
As shown in Fig.~\ref{fig:odl_flow_comp}(c), the feature vectors from the feature extractor (FE) are encoded and trained in the format of HV in the HDC classifier. The encoding used in \design is based on the binary random projection (RP) encoding~\cite{thomas2021theoretical} leading to the dramatic computing complexity reduction. RP uses a random binary base matrix $\mathbf{B} \in \{-1, +1\}^{D\times F}$ to encode the extracted $F$-dimensional features into $D$-dimensional HVs as:
\begin{equation}\label{eq:hd_enc}
   \mathbf{h} = \text{Encode}(\mathbf{x}) = \mathbf{B} \cdot \mathbf{x}.
\end{equation}
This ensures that the incoming data is projected on a hyper-dimensional space.

\subsubsection{Training and Inference:}
The efficiency of HDC mainly comes from its simplified training and inference process. The HDC training is based on the HV aggregation step: 
\begin{equation}\label{eq:hd_aggregate}
	\mathbf{C}_j = \sum_{i=1}^{k} \mathbf{h}_i^j,
\end{equation}
where $\mathbf{h}_i^j$ denotes the $i$-th encoded HV belonging to the $j$-th class with total $k$ encoded HVs for each class.  The final HDC model is constructed by storing the aggregated class HVs  for all classes, denoted as $\mathbf{C}_j$. 

The inference step compares the query HV $\mathbf{q}$ and finds the most similar class HV $\mathbf{C}_j$ via distance search: 
\begin{equation}\label{eq:hd_inference}
    \underset{j}{\arg\min}~\text{Distance}(\mathbf{q}, \mathbf{C}_j),
\end{equation}
Prediction is performed by comparing a query HV to the stored class HVs $\mathbf{C}_j$ and then selecting the class that yields the smallest distance (\eg, cosine or Hamming distance). A major advantage of HDC is its extremely lightweight training process, which requires only a single pass through the training data, avoids backpropagation or other gradient-based optimization, and relies on simple element-wise vector operations while still achieving competitive accuracy.

\section{\design ODL Algorithm}\label{sec:prposed_alg}


In this section, we present the \design algorithm, which comprises two key techniques for efficient feature extraction and classification, respectively.

\begin{figure}[t]
    \centering
    \includegraphics[width=0.85\linewidth]{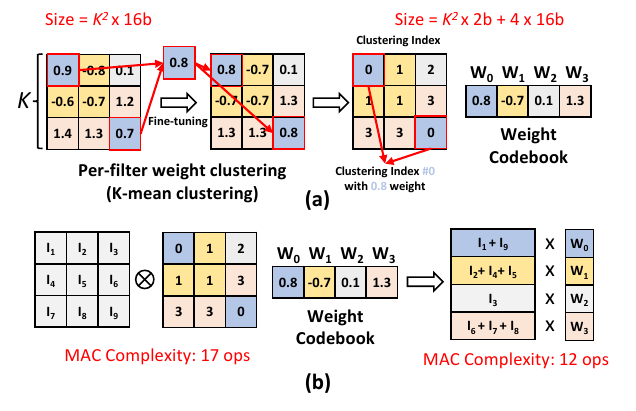}
    \caption{Weight clustering: (a) average weight clustering and index for each weight, (b) partial sum reuse based on common weight.}
    \label{fig:pnet_weight_clustering}
\end{figure}

\subsection{Low-complexity Feature Extraction with Weight Clustering}\label{subsec:fe_wc}

\design eliminates costly gradient computations ($C_{\text{GC}}$ and $C_{\text{BP}}$) by adopting a transfer learning-inspired approach~\cite{sun2019meta,tian2020rethinking,chowdhury2021few}, in which pre-trained CNN weights are frozen and reused for downstream tasks. Unlike existing ODL accelerators that retrain the entire CNN feature extractor, \design preserves the frozen extractor and focuses exclusively on training the lightweight HDC-based FSL classifier as shown in Fig.~\ref{fig:odl_flow_comp}(c).

With gradient computations removed from the CNN, feedforward inference becomes the primary bottleneck. 
To reduce the cost of feedforward, we implement a complexity-reduced FE that lowers ODL latency while preserving feature quality. The proposed FE exploits the parameter-efficient weight clustering technique~\cite{khaleghi2022patternet} to simultaneously reduce the required data movement and MAC operations. Fig.~\ref{fig:pnet_weight_clustering} shows two key steps: (a) weight clustering and (b) weight codebook sharing. After CNN pre-training, an additional step, weight clustering, is applied to generate the clustered weight as shown in Fig.~\ref{fig:pnet_weight_clustering}(a). Similar weights within $Ch_{\text{sub}}$ input channels are grouped into $N$ unique centroids using $K$-means clustering, \eg, 0.9 and 0.7 are grouped to be 0.8. Then, the centroids are converted into the corresponding weight index and codebook, \eg, the index is 0 for the weight 0.8. Note the weight index only requires the $\log_2{N}$-b for each weight, reducing the precision requirement. The $N$ centroid values, \eg, 0.8, are stored in the weight codebook. To enable high-precision feature extraction across diverse applications, FE adopts the bfloat16 (BF16) format, and each weight codebook has $N\times 16$-b data. 

The common patterns in clustered weights can be exploited to reduce inference complexity by reusing the partial sum as shown in Fig.~\ref{fig:pnet_weight_clustering}(b). The input activations in BF16 format with the same weight index are first accumulated in the accumulation buffer, \eg, $I_2$, $I_4$, and $I_5$ are summed. Then, the $N$ accumulated input activations are multiplied with $N$ weights from the codebook, and subsequently summed to obtain the final MAC output. The original $2 \times K^2 - 1$ ops are reduced to $K^2  + N - 1$ ops  in the proposed FE.


\begin{figure}[t]
    \centering
    \includegraphics[width=0.9\linewidth]{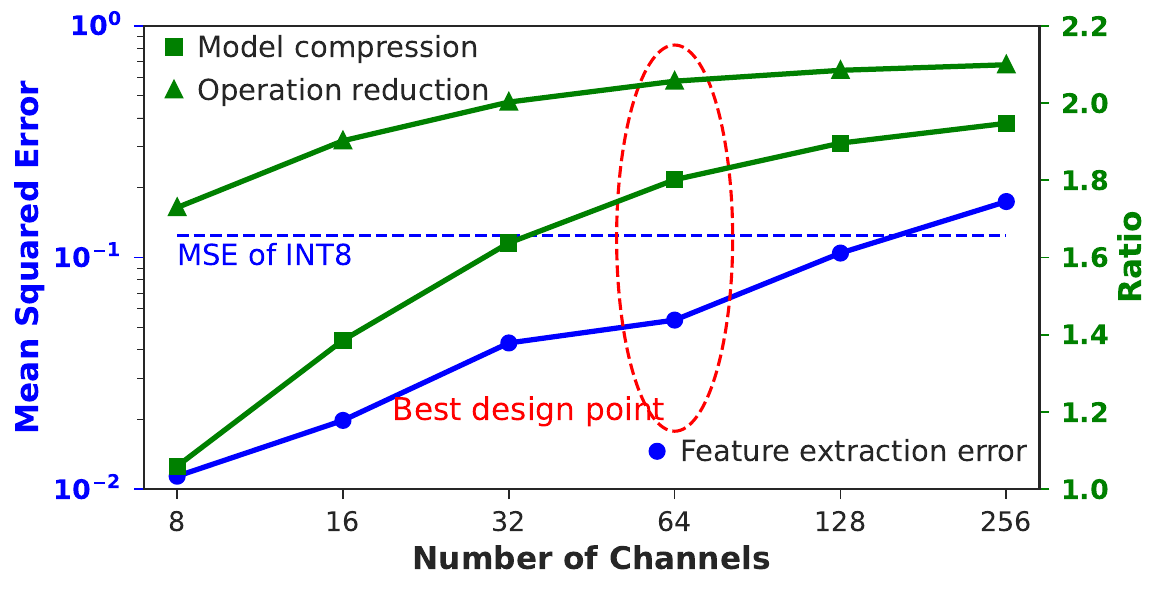}
    \caption{Feature extraction (FE) output error, model compression, and operation reduction of \design as compared to INT8-quantized ResNet-18 when varying $Ch_{\text{sub}}$ from 8 to 256.}
    \label{fig:rmse_mac_vs_cluster}
\end{figure}

Fig.~\ref{fig:rmse_mac_vs_cluster} illustrates the trends in FE error, model compression ratio, and operation reduction ratio with respect to $C_{\text{sub}}$, evaluated on ResNet-18\cite{he2016deep} ($K=3$) using the CIFAR-10 dataset~\cite{dataset_cifar10}, where the INT8-quantized model serves as the baseline.  
As $Ch_{\text{sub}}$ increases, the compression and operation reduction ratios improve as more input channels share the same codebook while the benefit saturates around at $2\times$ benefits for both.  
We choose $Ch_{\text{sub}}=64$ as the optimal design point because the corresponding FE error remains below INT8 model's mean squared error while $1.8\times$ memory saving and $2.1\times$ computational reduction are achieved.

\begin{figure}[t]
    \centering
    \includegraphics[width=0.9\linewidth]{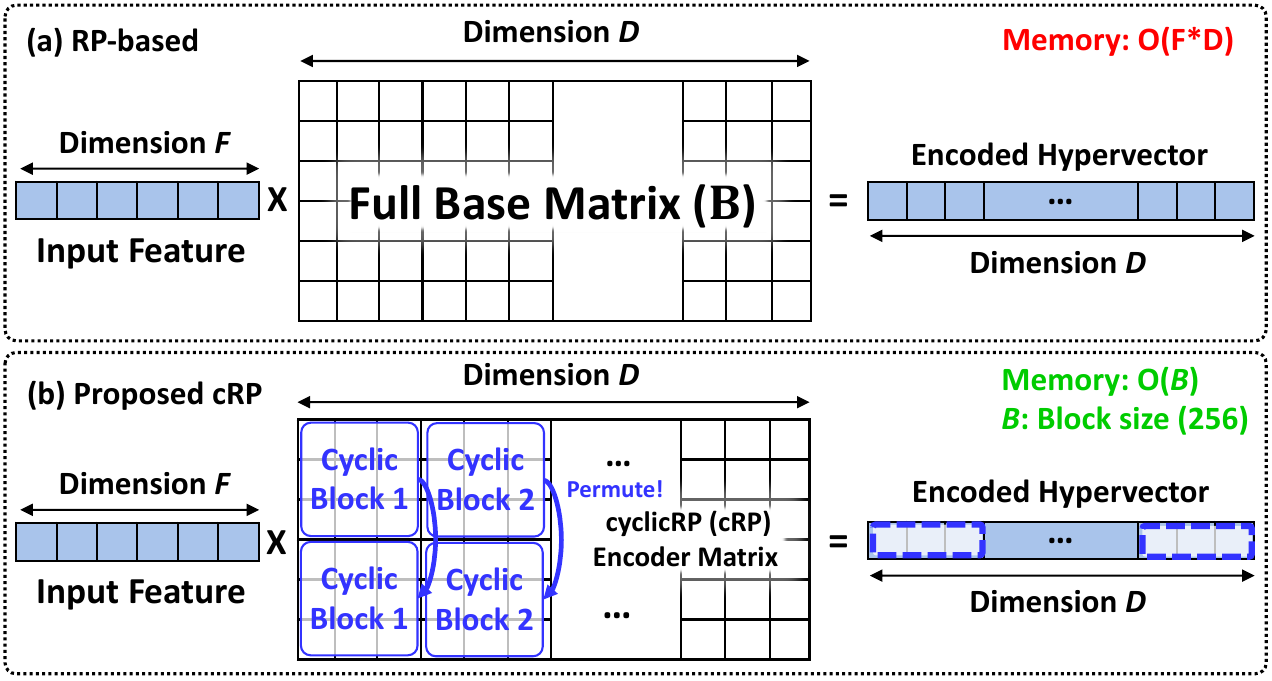}
    \caption{(a) Conventional RP-based HDC encoding, (b) proposed cRP-based HDC encoding.}
    \label{fig:cpr}
\end{figure}

\subsection{Single-Pass FSL with HDC}\label{prposed_alg:sp}
The additional source of inefficiency in existing ODL is the iteration $T_{\text{itr}}$ in (\ref{eq:odl_full_cost}) and (\ref{eq:odl_partial_cost}). \design leverages the single-pass training in HDC to skip the iterative training, thereby reducing ODL complexity and latency. The HDC learning process in \design includes following components.

\subsubsection{Memory-efficient Cyclic Random Projection (cRP) Encoding:}

To address the prohibitive memory footprint of the conventional Random Projection (RP) encoder, we implement a hardware-friendly cyclic Random Projection (cRP) encoder in \design. While conventional RP encoding~\cite{morris2021locality} in Fig.~\ref{fig:cpr}(a) achieves promising accuracy by multiplying a binary random base matrix $\mathbf{B}$ of size $F\times D$ with the input feature vector, this naive implementation requires substantial memory, \eg, 256KB for typical parameters ($F=512$, $D=4096$).
Our memory-efficient cRP encoder in Fig.~\ref{fig:cpr}(b) circumvents this limitation by generating RP weights on-the-fly instead of storing the entire base matrix. The approach decomposes the large matrix multiplication into smaller cyclic blocks of $B=16\times 16$ elements. A linear-feedback shift register (LFSR)-based pseudo-random number generator (PRNG) dynamically transforms the initial block for each cyclic block's position in Fig.~\ref{fig:cpr}, eliminating the need for explicit matrix.  This scheme effectively reduces the memory consumption of the base matrix from $\mathcal{O}(F\times D)$ to a constant $\mathcal{O}(B)$. 
 The detailed hardware implementation is in Section~\ref{subsec:crp}.

\subsubsection{Single-pass Training:} 
Unlike prior ODL chips that rely on gradient-based training, \design adopts a gradient-free, single-pass HDC learning paradigm, where the model is trained by processing each training sample only once. This single-pass nature removes iterative overhead not only in the HDC classification stage but also in feature extraction.
Fig.~\ref{fig:odl_flow_comp}(c) (and Fig.~\ref{fig:fsl_acc_comp} in Section\ref{sec:eval_measurement}) demonstrates that the gradient-free, single-pass \design achieves training convergence in a single iteration while delivering accuracy comparable to other gradient-based and kNN baselines. 
The complexity of \design ODL algorithm can be expressed as:
\begin{equation}\label{eq:odl_cost_proposed}
    \text{Cost}_{\text{\design}} \approx N_{\text{sample}} \times (\text{Cost}_{\text{FP}} + \text{Cost}_{\text{HDC}}),
\end{equation}
where the iteration $T_{\text{itr}}$ is eliminated, unlike full and partial FT cases,  while the gradient operations' costs ($C_{\text{GC}}$, $C_{\text{BP}}$, and $C_{\text{WU}}$ in (\ref{eq:odl_full_cost}) and (\ref{eq:odl_partial_cost})) are replaced by HDC training cost $C_{\text{HDC}}$. 
The lightweight nature of the proposed FE  in \design further reduces the required ODL complexity.
Fig.~3(b) shows that FSL-HDnn achieves the best tradeoff between accuracy and training complexity among all baselines, making it well-suited for on-device applications.


\begin{figure}[t]
    \centering
    \includegraphics[width=0.87\linewidth]{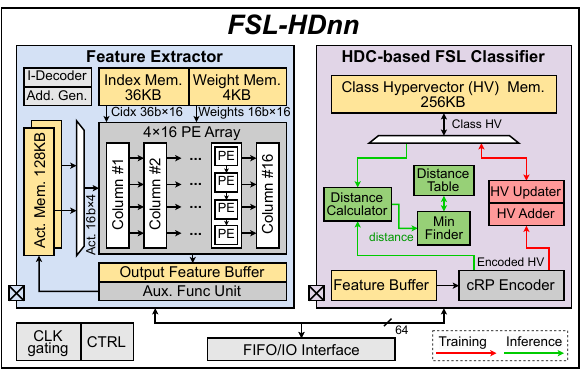}
    \caption{Proposed \design chip architecture.}
    \label{fig:chip_overall}
\end{figure}

\section{\design Chip Architecture}\label{sec:proposed_arch}

Fig.~\ref{fig:chip_overall} illustrates the overall architecture of \design, which consists of a FE module, a HDC-based FSL classifier, IO FIFO interface, and other peripheral modules (clock gating and controller). 
Either the features extracted by FE or the raw input data can serve as the input to the FSL classifier.
The details for each module are described below. 

\subsection{Weight Clustering-based Feature Extractor}\label{sec:FE}

\subsubsection{Architecture and Dataflow:}
Fig.~\ref{fig:chip_overall} illustrates the overall architecture of the proposed FE module that leverages the weight clustering mechanism in Section~\ref{subsec:fe_wc}. The architecture has an 8-bank, 128KB activation memory with double buffering to minimize IO stalls during activation loading. Weight storage is partitioned between a 16-bank, 36KB index memory containing $\log_2{N}$-bit weight indices and a 16-bank, 4KB weight memory storing BF16 codebooks.
The core computation occurs in a $4\times 16$ PE array employing a codebook-stationary dataflow for optimal efficiency. The PE in each column is dedicated to a specific output channel while 
the four rows operate on different activations in the same column of the input feature map, taken from consecutive input rows, to produce four output pixels in consecutive rows in parallel. 
Note that all the PEs within a column of PE array process the same position within their convolution windows and the same input channel; this means they share the same weight index and value from the codebook. Therefore, the weight indices and their corresponding codebook values are broadcast to the  PEs within each column.

In this process, all input pixels within the convolution window are streamed one by one to the PEs for a given channel, and the computation then proceeds to the next channel. After all channels in $Ch_{\text{sub}}$ are processed, the codebook for the next $Ch_{\text{sub}}$ is loaded. Once computations for the current window position are complete, the window shifts to generate the next set of output pixels. This procedure is repeated until all output channels are covered.


\begin{figure}[t]
    \centering
    \includegraphics[width=0.95\linewidth]{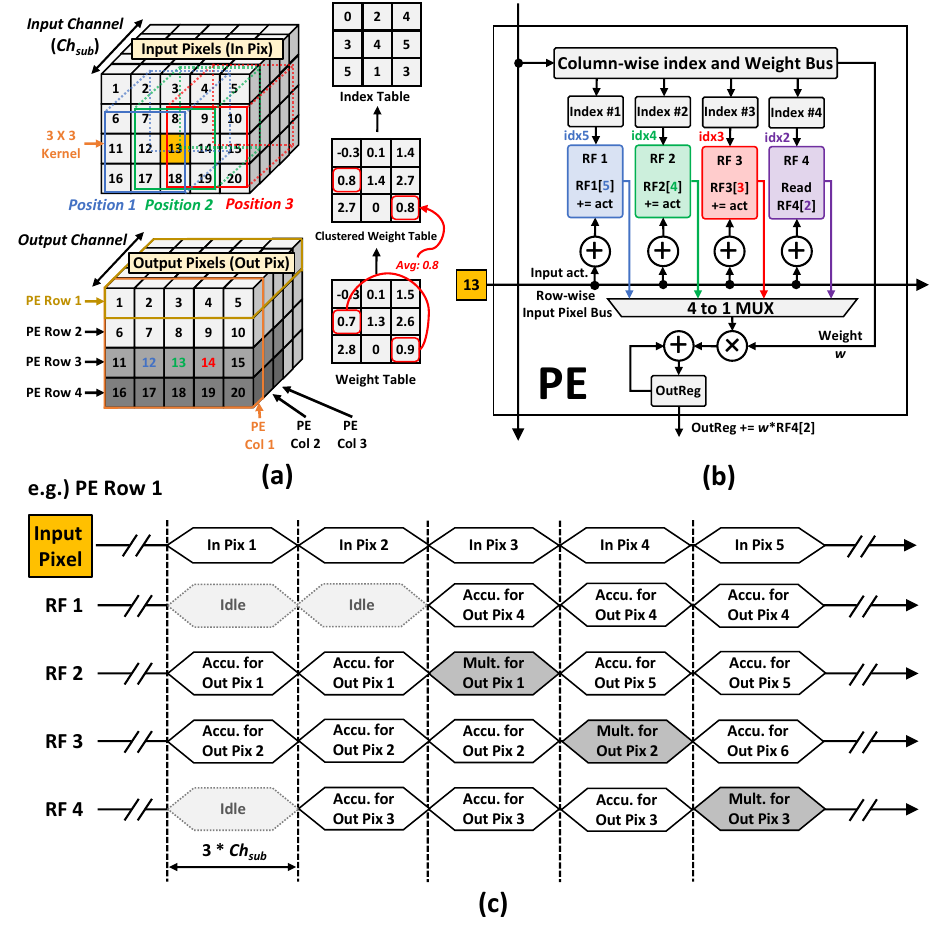}
    \caption{(a) Feature extraction with weight clustering, (b) Feature extractor
processing element (PE), (c) PE timing diagram.}
    \label{fig:fe_hardware}
\end{figure}

\subsubsection{PE Design:}
Fig.~\ref{fig:fe_hardware}(b) shows the PE design that contains four RFs, four adders, and one MAC unit. As shown in Fig.~\ref{fig:pnet_weight_clustering}(b), the weight clustering-based CONV consists of two steps: 1. accumulation for input activations with the same index, and 2. MAC operation for codebook values. The PE architecture is optimized for 3$\times 3$ kernels and maximizes hardware utilization by processing the accumulation and MAC operation steps in parallel. Specifically, three of these RFs are responsible for processing three horizontally consecutive output pixels using the same CONV filter, while the remaining RF is used to perform MAC operations. 
The three input weight indices are regarded as the RF addresses to fetch the corresponding partial sum from three RFs. As shown in Fig.~\ref{fig:fe_hardware}(a) and (b), if RF1 receives the index “5”, the input, 13 in the yellow box, is accumulated in the fifth element of the RF to generate the partial sum when the convolution window is in the position 1.

These operations are performed in the three RFs to generate three horizontally adjacent output pixels in parallel, processing the convolution windows at three positions of the input feature map, shown in blue, green, and red.
While three RFs perform accumulation, the accumulated input from the remaining RF (that has completed accumulation) is multiplied with BF16 weight values from the weight bus for that pixel. In Fig.~\ref{fig:fe_hardware}(c), the time-wise overlapped execution ensures that the MAC operation of the previous output pixel will not stall the accumulation of new inputs, maintaining a high PE utilization.

\subsection{HDC-based FSL Classifier}\label{subsec:crp}

\begin{figure}[t]
    \centering
    \includegraphics[width=0.8\linewidth]{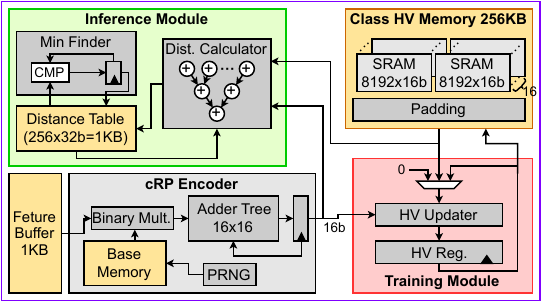}
    \caption{Hardware architecture of HDC-based FSL classifier.}
    \label{fig:hdc_arch}
\end{figure} 

\subsubsection{Architecture Design:}
Fig.~\ref{fig:hdc_arch} illustrates the hardware architecture of \hd-based FSL classifier composed of the cyclic Random Projection (cRP) encoder, inference module, and training module. 
The cRP encoder caches the incoming $F$-dimensional feature in the feature buffer and then encodes the feature into $D$-dimensional HV. The distance calculation module is used for inference and computes the distance between query HV and class HVs. The training module is responsible for generating and updating class HVs.



\subsubsection{cRP Encoder:} \label{sec:crp}

\begin{figure}[t]
    \centering
    \includegraphics[width=0.85\linewidth]{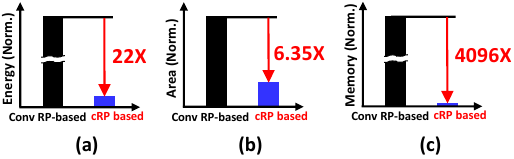}
    \caption{(a) Energy efficiency, (b) area efficiency, and (c) memory efficiency improvements using cRP-based encoding.}
    \label{fig:fsl_result}
\end{figure}

In the cRP encoder, a $16 \times 16$ binary RP weight block is generated by the PRNG in a cycle and cached in the base memory. The PRNG consists of 16 LFSRs~\cite{oommen2018study}, each producing a 16-bit output, resulting in a total of 256 bits per cyclic block. By storing only the initial block generated from a random seed, the entire RP matrix can be reconstructed on demand by repeatedly advancing the LFSRs through their deterministic shift-and-feedback cycles, without explicitly storing the full matrix. 
The encoding process is performed for each segment sequentially for $\frac{D\times F}{B}$ cycles where $B=16\times 16$ elements constitute one cyclic block. (Fig.~\ref{fig:cpr}(b)).
Fig.~\ref{fig:fsl_result} shows that the cRP encoder consumes 22$\times$ less energy, 6.35$\times$ less area, and 512 - 4096$\times$ weight memory compared to the conventional RP encoder~\cite{morris2021locality}.

The generated block is then segment-wise binary-multiplied with the corresponding 16-dimensional feature segment to produce the $16\times 16$ multiplication results as shown in Fig.~\ref{fig:cpr}. The 16 adder trees, each accepting 16 inputs, perform reduction on the multiplication results and produce the 16-dimensional segmented HV.




\subsubsection{Inference Module:}
The distance calculation module performs  HDC inference by computing the distance between the encoded HV from the cRP encoder and the class HVs stored in class memory. In each cycle, a 256-bit ($16 \times 16$) HV segment is fetched from the class memory.  The 256~KB class memory, organized as 16 SRAM banks, provides flexible storage capacity, supporting up to 32 class HVs at 16-bit precision or 128 class HVs at 4-bit precision (both with $D=4096$). This capacity is also sufficient to support the early exit inference capabilities described in Section~\ref{sec:proposed_opt}. During inference, each element of the encoded HV is subtracted from the corresponding element of the class HV, and the absolute differences of each element are accumulated to compute the distance. The class HV with the minimum distance to the input HV determines the final classification result. To reduce power consumption, unused SRAM banks are gated off. 

\subsubsection{Training Module:}
The training module contains a class memory and an HV updating module. HVs are processed with configurable precision from 1-bit to 16-bit integers, accommodating diverse task requirements. During training, a 256-bit ($16\times16$) HV segment is fetched from class memory and loaded into the HV register each cycle. The HV updater, featuring 16 parallel adders, each being a 16-term adder with flexible 1- to 16-bit input precision, executes the aggregation-based HDC training defined in (\ref{eq:hd_aggregate}) for each segment.


\section{Optimizations for Low-latency Few-shot ODL}\label{sec:proposed_opt}

Low-latency inference and training are critical for real-time adaptation and decision-making on device. In this section, we propose various optimization strategies to achieve low-latency ODL while improving energy efficiency.

\subsection{Inference with Early Exit and Branch Feature Extraction}

\begin{figure}[t]
    \centering
    \includegraphics[width=0.85\linewidth]{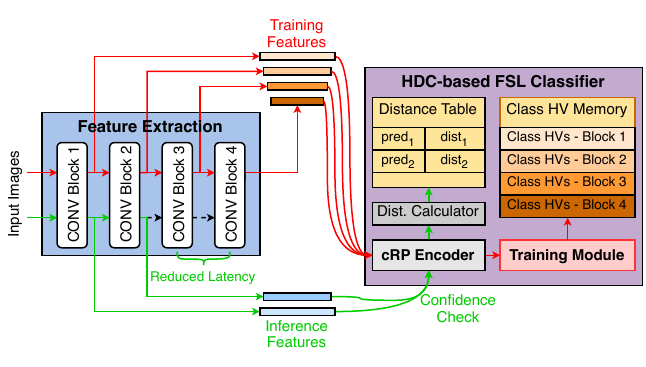}
    \caption{Early exit strategy in \design with branch feature extraction and HDC-based confidence check. Each CONV block of ResNet18 (Fig.~\ref{fig:early_exit}) contains four CONV layers. }
    \label{fig:early_exit}
\end{figure}

Our analysis reveals that not all samples require exhaustive feature extraction, \ie, simpler samples can achieve accurate predictions using only shallow feature extractions as in Fig.~\ref{fig:early_exit}.
To boost the inference efficiency by leveraging this nature, \design adopts the early exit (EE) mechanism (Fig.~\ref{fig:early_exit}). 
Early exit~\cite{teerapittayanon2016branchynet,jo2023locoexnet,rahmath2024early,ilhan2024adaptive} is an effective strategy for reducing inference complexity by adaptively terminating computation at intermediate layers when predictions are sufficiently confident. However, enabling this capability with conventional early exit requires branch feature extraction and additional training cost for branch heads, which limits its practicality for ODL. In this work, we overcome these limitations by combining the early exit mechanism with lightweight HDC for better efficiency.

\noindent \textbf{$\bullet$ Training:} The EE training follows the single-pass paradigm in Section~\ref{sec:prposed_alg}. During training, the auxiliary function unit (AFU) in the feature extractor (Fig.~\ref{fig:chip_overall}) computes the average pooling  of  each CONV block's output for the data with the same label, generating feature vectors of varying dimensions. These features are then uniformly encoded into $D$-dimensional binary HVs by the cRP encoder in the HDC-based FSL classifier. As shown in Fig.~\ref{fig:early_exit}, each input image produces four feature vectors (one per CONV block in ResNet-18), which are encoded and stored as corresponding class HVs in memory. The primary overhead of EE training is the increased memory consumption for storing branch-specific class HVs. For a $C$-way FSL task with $B$-bit class HVs, the total memory footprint is $4C\cdot D\cdot B$ bits. The 256~KB class HV memory accommodates up to 32-way FSL tasks with $D=4096$ and 4-b class HVs, providing sufficient capacity for practical applications.

\noindent \textbf{$\bullet$ Inference:} 
Fig.~\ref{fig:early_exit} illustrates how features from the FE are progressively evaluated by the HDC classifier during inference. As each CONV block generates its output features, they are immediately sent to the HDC-based FSL classifier for HV encoding and confidence check. The HV generated by an intermediate block is compared with the corresponding block’s class HVs. For example, the feature captured from the third CONV layer is compared with the class HVs generated from the third CONV layer. We introduce a lightweight confidence check mechanism that requires no additional hardware: the FSL classifier terminates inference when predictions remain consistent across $E_c$ consecutive CONV blocks, starting from the $E_s$-th CONV block. 
The distance table in the FSL classifier stores the prediction results from previous CONV blocks, enabling the distance calculation module to verify prediction consistency based on the configured ($E_{s}$, $E_{c}$) parameters. These parameters control the tradeoff between FSL accuracy and inference efficiency: larger $E_s$ and $E_c$ values preserve accuracy at the expense of reduced latency savings, while smaller values achieves aggressive speedup with modest accuracy degradation (see results in Section~\ref{subsec:e2e_eval}).

\begin{figure}[t]
    \centering
    \includegraphics[width=0.85\linewidth]{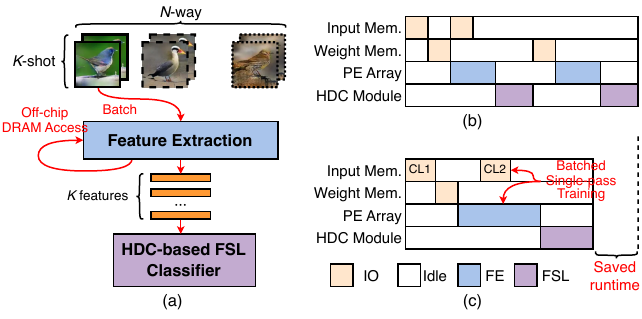}
    \caption{(a) Batched single-pass training. Timing diagrams  (b)  without batching, (c) with batching. (CL = class)}
    \label{fig:batch_training}
\end{figure}

\subsection{Batched Single-pass Training}

While the double-buffered input memory eliminates the impact of activation loading on feature extraction runtime, the PE array still experiences stalls due to weight index and codebook loading operations, as shown in Fig.~\ref{fig:batch_training}(b). To address this bottleneck and improve hardware utilization, we introduce batched single-pass training in Fig.~\ref{fig:batch_training}(a). While the batched processing has been widely exploited for the efficient training, the batched processing provide even greater synergy with HDC.  This approach exploits the observation that: for FSL tasks, training HVs belonging to the same class can be aggregated in a single pass.
Batched single-pass training enhances weight reuse by processing multiple samples together, significantly reducing  stalls from weight loading. In an $N$-way $K$-shot FSL task, the FE extracts all $K$ features for each class, which are then aggregated and encoded only once to produced the trained HVs in the HDC module. This strategy minimizes weight loading operations, and maximizes PE array utilization, as depicted in Fig.~\ref{fig:batch_training}(c) along with minimal HD encoding overhead. Section~\ref{subsec:e2e_eval} presents detailed experimental results validating the effectiveness of this batched training approach.




\section{Results and Analysis}\label{sec:eval_measurement}

\subsection{\design Chip Implementation}
\begin{figure}[t]
    \centering
    \begin{subfigure}[b]{1\linewidth}
        \centering
        \includegraphics[width=\linewidth]{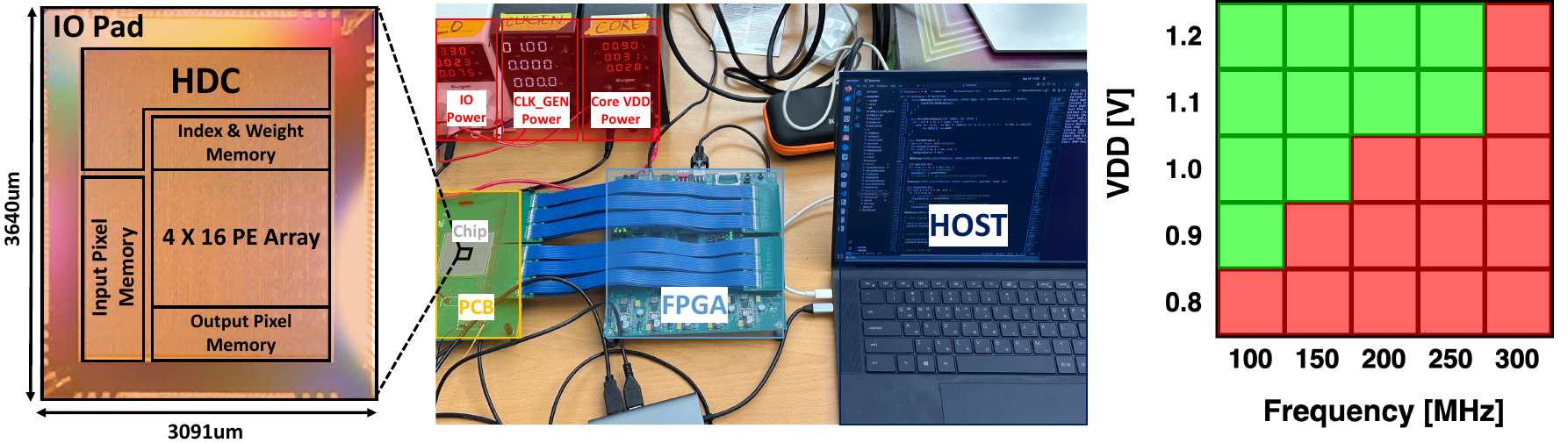}
        \caption{}
        \vspace{0.1cm}
    \end{subfigure}
    \hfill
    \begin{subfigure}[b]{0.59\linewidth}
        \scriptsize
        \centering
        \renewcommand{\arraystretch}{1.1}
        \resizebox{\linewidth}{!}{
        \begin{tabular}{|>{\columncolor{myGray}}c|c|}
            \hline
            Technology & 40~nm CMOS \\
            \hline
            Die Area & 3.64~mm$\times$3.09~mm \\
            \hline
            On-chip SRAM & 424~KB \\
            \hline
            & BF16 for CNN \\
            \multirow{-2}{*}{Precision} & INT1-16 for HDC \\
            \hline
            Supply Voltage & 0.9-1.2~V \\
            \hline
            Frequency & 100-250~MHz \\
            \hline
            Model & CNN+HDC \\
            \hline
            FSL Feature Dimension ($F$) & 16-1024 \\
            \hline
            FSL Classes ($N$) & 2-128 \\
            \hline
            HDC Dimension ($D$) & 1024-8192 \\
            \hline
            Power & 59-305~mW \\
            \hline
            Training & 2.9~TOPS/W @ 0.9~V, 100~MHz \\
            Energy Efficiency & 1.4~TOPS/W @ 1.2~V, 250~MHz \\
            \hline
        \end{tabular}
        }
        \caption{}
    \end{subfigure}
    \caption{\design chip (a) micrograph, verification setup,  shmoo plot, and (b) specifications.}
    \label{fig:chip_mg_shmoo_spec}
\end{figure}

The \design chip was fabricated using  40~nm CMOS technology. Fig.~\ref{fig:chip_mg_shmoo_spec}(a) shows the chip micrograph,  its verification setup, and shmoo plot. The chip with an area of 11.3~mm$^2$ is packaged using Pin Grid Array (PGA)  and integrated onto a test PCB. The Xilinx Zynq-7000 FPGA is connected to the test board via the mezzanine connector. 
Fig.~\ref{fig:chip_mg_shmoo_spec}(b) provides the detailed chip specifications. The \design chip with 424~KB on-device memory implements BF16 precision for feature extraction and INT1-16 for HDC-based FSL classification. A wide range of dimensions are supported: the feature dimension ($F$) ranges from 16 to 1024 while the HDC dimension ($D$) ranges 1024 to 8192 with a maximum of 128 classes.

\subsection{Measurement Results}

\begin{figure}[t]
    \centering
    \begin{subfigure}[b]{0.47\linewidth}
        \centering
        \includegraphics[width=\linewidth]{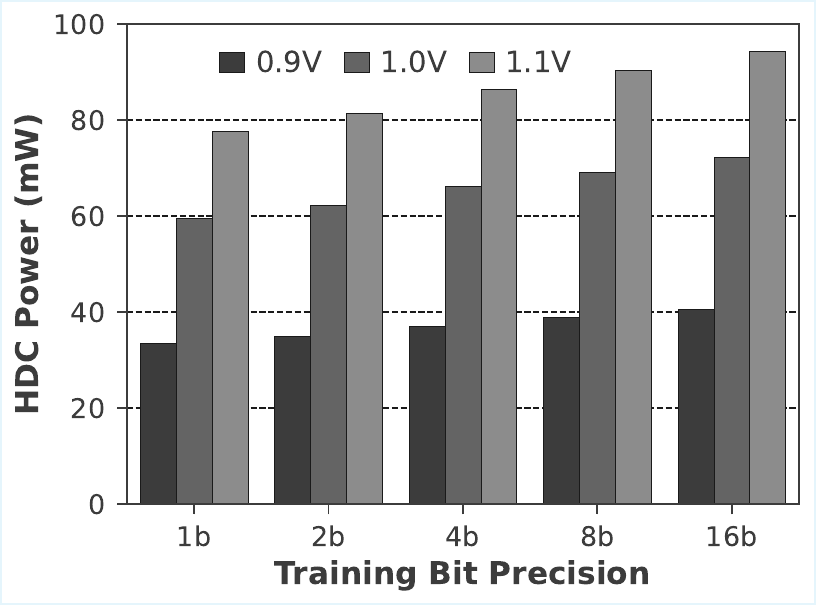} 
        \caption{HDC}
    \end{subfigure}
    \begin{subfigure}[b]{0.28\linewidth}
        \centering
        \includegraphics[width=\linewidth]{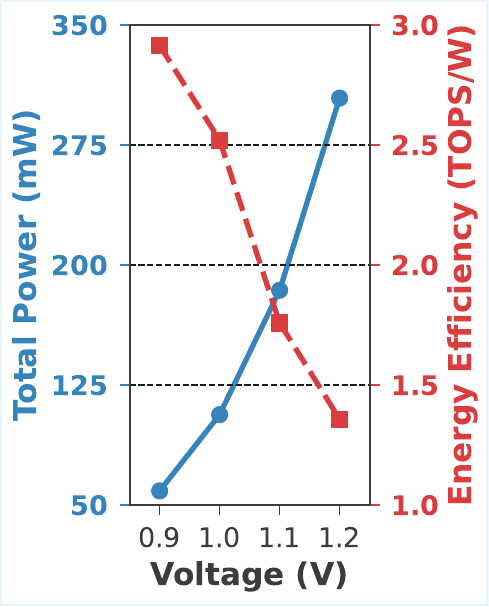}
        \caption{Chip}
    \end{subfigure}
    \caption{Measured power consumption of (a) the training  of HDC-based FSL classifier module in Fig.~\ref{fig:chip_overall}, and (b) overall \design chip.}
    \label{fig:power}
\end{figure}

\design is extensively evaluated on three few-shot classification datasets (CIFAR100~\cite{dataset_cifar100}, Flower102~\cite{dataset_flower}, and Traffic-sign~\cite{dataset_traffic}). All input images are resized to $224\times 224$. The ResNet18~\cite{he2016deep} model pretrained on ImageNet~\cite{deng2009imagenet} is adopted as the feature extractor and the feature output is quantized to 4-b. The default HDC dimension is $D=4096$.
The measured results show that the operating frequency of 250~MHz is achieved with 1.2~V supply voltage. Fig.~\ref{fig:power}(a) shows the measured training power of the HDC-based FSL classifier given in Fig.~\ref{fig:chip_overall} with different precisions and voltages. The HDC module consumes 21\% more power when bit precision increases from 1-b to 16-b, mainly due to the higher power demand of distance computations and more memory accesses. Fig.~\ref{fig:power}(b) shows the curve of total power and energy efficiency w.r.t supply voltages. \design chip dissipates 59~mW at 100~MHz and 0.9~V to 305~mW at 250~MHz and 1.2~V. The achieved energy efficiency for the training of FSL is 1.4 to 2.9~TOPS/W.


\subsection{Evaluation for End-to-end FSL}\label{subsec:e2e_eval}

\subsubsection{FSL Accuracy:}

\begin{figure}[t]
    \centering
    \includegraphics[width=0.9\linewidth]{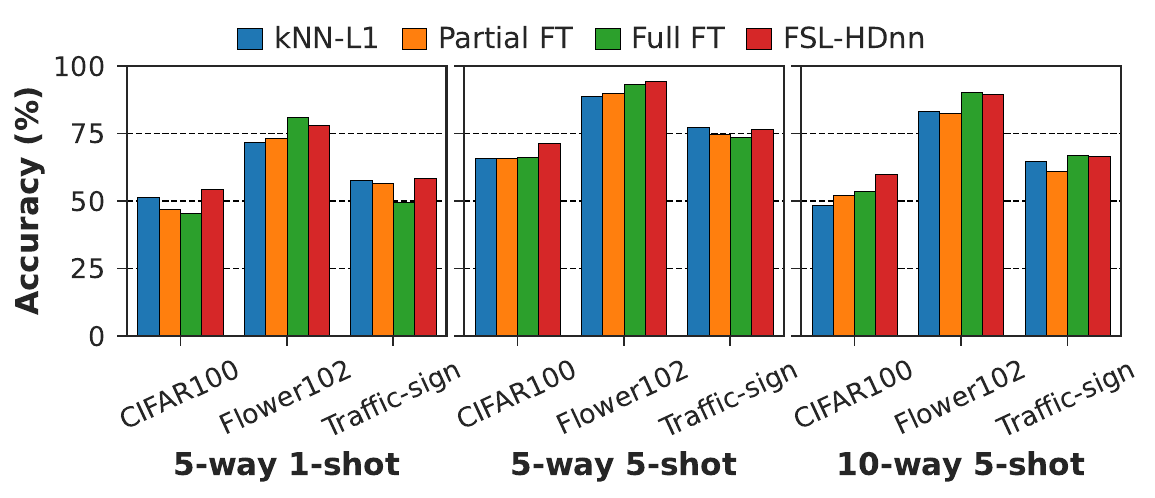} 
    \caption{FSL accuracy comparison. Full FT and partial FT are trained for five epochs and 15 epochs, respectively. kNN-L1 and \design are based on single-pass training.}
    \label{fig:fsl_acc_comp}
\end{figure}

Fig.~\ref{fig:fsl_acc_comp} depicts the FSL accuracy of \design compared to kNN-L1~\cite{li2021sapiens,chowdhury2021few} and FT-based FSL classifiers~\cite{han2021dflnpu,park2021neural,prabhu2022chimera,wang2022trainer,venkataramanaiah202328,qian20244,cai2020tinytl} under various learning settings. The results highlight \design's ability to match FT-based accuracy (\eg, 94.1\% vs. 94.5\% on Flower102) while surpassing kNN by 4.9\% on average, with the large margin (6.3\%) on Traffic-sign. This result highlights the effectiveness of HDC-based FSL, where simplistic kNN distance metrics fall short. Notably, \design achieves this performance using single-pass gradient-free training, reducing the number of computing operations by $21\times$ compared to FT-based methods. The consistent gains across datasets underscore the robustness of the proposed gradient-free ODL algorithm, making it well-suited for edge devices where both accuracy and efficiency are critical.

\begin{table*}[t]
    \centering
    \caption{Comparison with state-of-the-art ODL accelerators.}
    \label{tab:comparison}
    \scriptsize
    \renewcommand{\arraystretch}{1.05}
    \resizebox{0.88\linewidth}{!}{
    \begin{threeparttable}
    \begin{tabular}{|>{\columncolor{myGray}}c|c|c|c|c|c|c|c|}
        \hline
        \rowcolor{myGray}
        & \textbf{DF-LNPU} & & \textbf{CHIMERA} & \textbf{Trainer} & & & \textbf{\design} \\
        \rowcolor{myGray}
        & \textbf{JSSC'21~\cite{han2021dflnpu}} & \multirow{-2}{*}{\textbf{JSSC'22~\cite{park2021neural}}}  & \textbf{JSSC'22~\cite{prabhu2022chimera}} & \textbf{JSSC'22~\cite{wang2022trainer}}& \multirow{-2}{*}{\textbf{JSSC'23~\cite{venkataramanaiah202328}}} & \multirow{-2}{*}{\textbf{JSSC'24~\cite{qian20244}}} & \textbf{(This work)} \\ 
        \hline
        \textbf{Technology} & 65~nm  & 40~nm & 40~nm & 28~nm & 28~nm & 28~nm & 40~nm\\
        \hline
        \textbf{Implementation} & Digital  & Digital & Digital-ReRAM & Digital & Digital & Digital & Digital \\
        \hline
         \textbf{Die Area (mm$^2$)} & 5.36 & 6.25 & 29.2 & 20.9 & 16.4 & 2 & 11.3 \\
        \hline
        \textbf{Frequency (MHz)} & 25-200 & 20-180 & 200 & 40-440 & 75-340 & 20-200 & 100-250 \\
        \hline
        \textbf{Voltage (V)} & 0.7-1.1 & 0.75-1.1 & 1.1 & 0.56-1.0 & 0.6-1.1 & 0.43-0.9 & 0.9-1.2\\
        \hline
        \textbf{On-chip Memory (KB)} & 168 & 293 & 2,560 & 634 & 1,280 & 64 & 424 \\
        \hline
        \textbf{Power (mW)} & 17.9-252.4 & 13.1-230 & 135 & 23-363 & 51.1-623.7 & 0.8-18 & 59-305 \\
        \hline
        \textbf{Precision} & INT16 & FP8  & INT8 & FP8/16 & INT8 & INT8 & BF16/INT1-16 \\
        \hline
        \textbf{Training Algorithm$^a$} & DFA BP + Partial FT & LP BP + Full FT & LR BP + Partial FT & Sparse BP + Full FT & Sparse BP + Full FT & Sparse BP + Full FT & HDC-based FSL \\
        \hline
        & & & & & & & 93.4 @ Flower102$^{b,d}$ \\
        & & & & & & & 78.3 @ TrafficSign$^{b,d}$ \\
        \multirow{-3}{*}{\textbf{Accuracy (\%)}} & \multirow{-3}{*}{42.0 @ Obj. Track$^b$} & \multirow{-3}{*}{69.0 @ ImageNet$^c$} & \multirow{-3}{*}{69.3 @ Flower102$^b$} & \multirow{-3}{*}{70.7 @ CUB-200$^b$} & \multirow{-3}{*}{94.3 @ CIFAR-10$^c$} & \multirow{-3}{*}{96.1 @ AntBee$^b$} &  72.5 @ CIFAR-100$^{b,d}$ \\
        \hline
        \textbf{Throughput (GOPS)} & 155.2 & 567 & 920 & 450 (FP16) & 560 & 38.4 & 197\\
        \hline
        \textbf{Energy Eff. (TOPS/W)}$^e$ & 0.8-1.5 & 1.6 @ ResNet18 & 2.2 (peak) @ ResNet18 & 0.9-1.6 & 4.1 @ ResNet20 & 1.6-3.6 & 1.4-2.9 @ ResNet18 \\
        \hline
        \textbf{Hardware Eff. (GOPS/mm$^2$)$^e$} & 78.8 & 90.7 & 31.5 & 10.1 & 15.9 & 9.0 & 16.7 \\
        \hline
        \textbf{FSL Training Latency$^f$} & \multirow{2}{*}{308 (8.9$\times$)} & \multirow{2}{*}{184 (5.3$\times$)} & \multirow{2}{*}{795 (23.0$\times$)} & \multirow{2}{*}{706 (20.4$\times$)} & \multirow{2}{*}{200 (5.8$\times$)} & \multirow{2}{*}{7,927 (229.1$\times$)}  & \multirow{2}{*}{35 (1.0$\times$)} \\
        \textbf{(ms/Image)} & & & & & & & \\
        \hline
        \textbf{FSL Training Energy$^f$} & \multirow{2}{*}{39 (6.5$\times$)} & \multirow{2}{*}{33 (5.6$\times$)} & \multirow{2}{*}{91 (15.2$\times$)} & \multirow{2}{*}{36 (6.1$\times$)} & \multirow{2}{*}{125 (20.9$\times$)} & \multirow{2}{*}{12 (2.0$\times$)}  & \multirow{2}{*}{6 (1.0$\times$)} \\
        \textbf{(mJ/Image)} & & & & & & & \\
        \hline
    \end{tabular}
    \begin{tablenotes}
        \item[a] DFA = Direct feedback alignment, LP = Low-precision, LR = Low-rank.
        \item[b] Transfer learning accuracy.
        \item[c] Accuracy by training from scratch.
        \item[d] 5-way 5-shot FSL accuracy.
        \\
        \item[e] Scaled to 40~nm~\cite{sarangi2021deepscaletool}. 
        \item[f] Estimated using 10-way 5-shot task and $224\times224$ image @ ResNet-18. Five training epochs are used for other baselines.
    \end{tablenotes}
    \vspace{-1.2cm}
    \end{threeparttable}
    }
\end{table*}

\begin{figure}[t]
    \centering
    \includegraphics[width=0.85\linewidth]{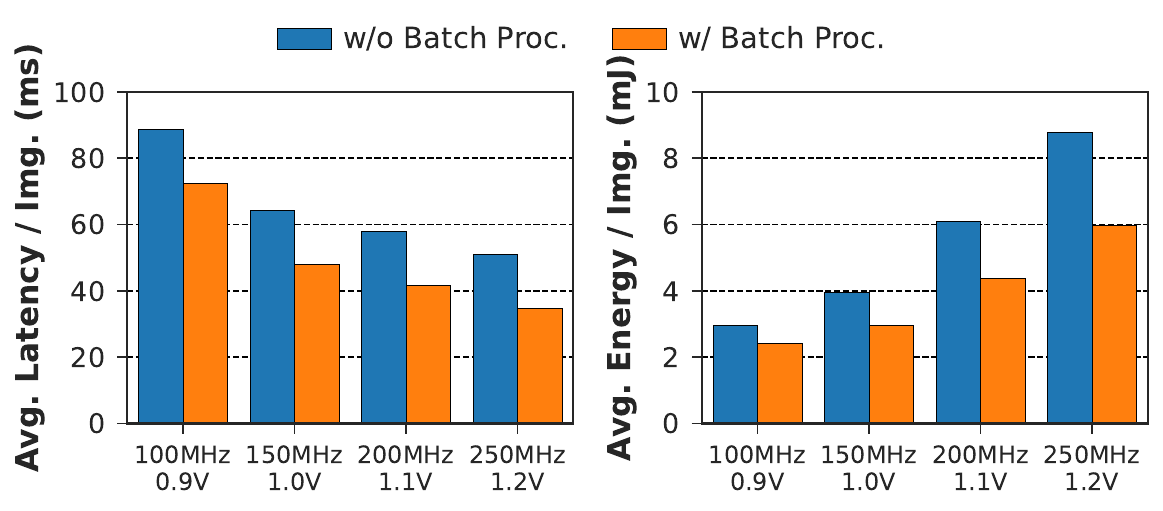} 
    \caption{Average latency and energy consumption per image of FSL training with and without batch processing.} 
    \label{fig:perf_training}
\end{figure} 

\subsubsection{Training:}
Fig.~\ref{fig:perf_training} illustrates the gains of using the proposed batched single-pass training. The batched single-pass training achieves 18\% to 32\% per-image latency and energy savings compared to non-batched processing. By grouping multiple samples, the architecture minimizes stall cycles due to weight loading and maximizes PE utilization via parallel computation. The speedup and energy gains are more pronounced in high-frequency regimes, where external I/O stalls to access off-chip DRAM consume a relatively higher runtime cycles.
Notably, the batched training retains the \design’s single-pass learning advantage by eliminating iterative gradient updates, achieving  6~mJ/image training energy efficiency.

\begin{figure}[t]
    \centering
    \begin{subfigure}[b]{\linewidth}
        \centering
        \includegraphics[width=0.85\linewidth]{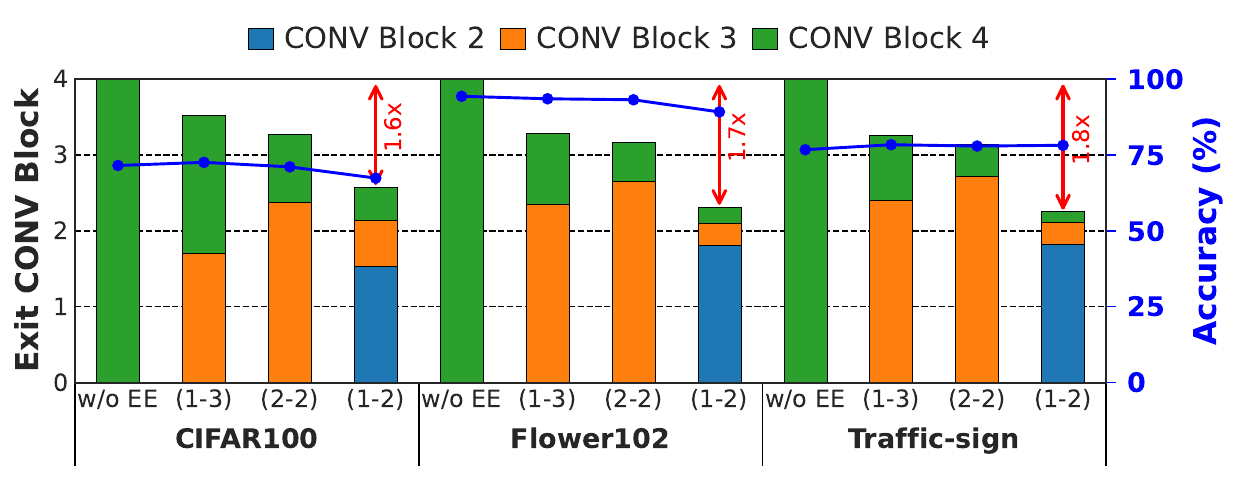}
        \vspace*{-3mm}
        \caption{5-way 5-shot}
        \vspace{-1mm}
    \end{subfigure}
    \begin{subfigure}[b]{\linewidth}
        \centering
        \includegraphics[width=0.85\linewidth]{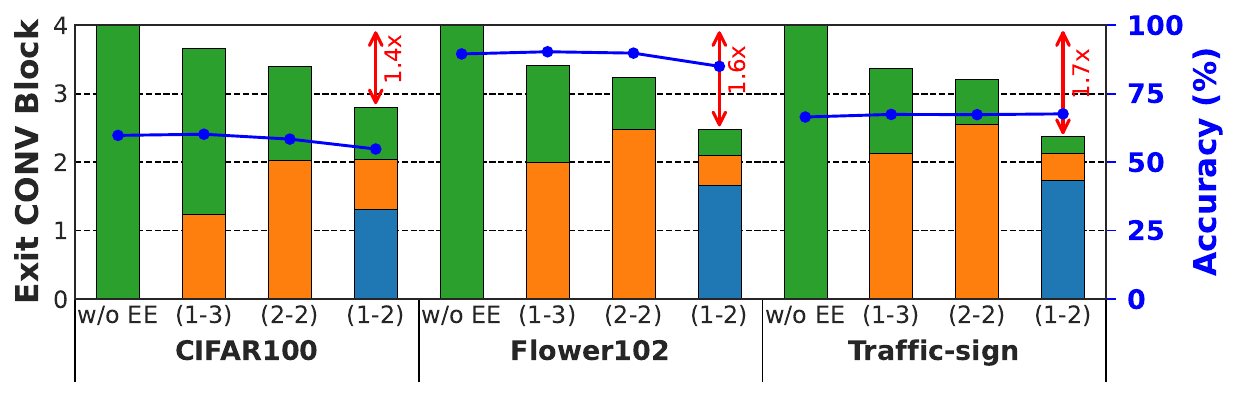}
        \vspace*{-3mm}
        \caption{10-way 5-shot}
    \end{subfigure}
    \caption{Average existing layers and FSL accuracy using different early-exit configurations ($E_s$-$E_c$). Each CONV block of ResNet18 (Fig.~\ref{fig:early_exit}) contains four CONV layers. \vspace{-2mm}}
    \label{fig:acc_ee_tradeoff}
\end{figure}

\subsubsection{Inference:}
Fig.~\ref{fig:acc_ee_tradeoff} shows the impact of early exit (EE) parameters (starting block \(E_s\) and consecutive confirmations \(E_c\)) on the FSL accuracy and exiting depth. Smaller \(E_s\) (earlier exits) and \(E_c\) (fewer confirmations) reduce the average layers processed by up to $45\%$ (\eg, \(E_s=1\), \(E_c=2\)) but incur a moderate accuracy drop ($\approx3.5\%$ on CIFAR-100). 
Conversely, stricter EE configurations (\(E_s=1\), \(E_c=3\)) retain near-optimal accuracy while still skipping 15\% to 20\% of CONV layers. The optimal balance is achieved at \(E_s=2\), \(E_c=2\), which bypasses 20\% to 25\% of layers with only $<1\%$ accuracy loss. These results demonstrate that the EE scheme adaptively lowers inference cost according to the classification difficulty of each input, thereby achieving efficient and low-latency inference.

\begin{figure}[t]
    \centering
    \includegraphics[width=0.85\linewidth]{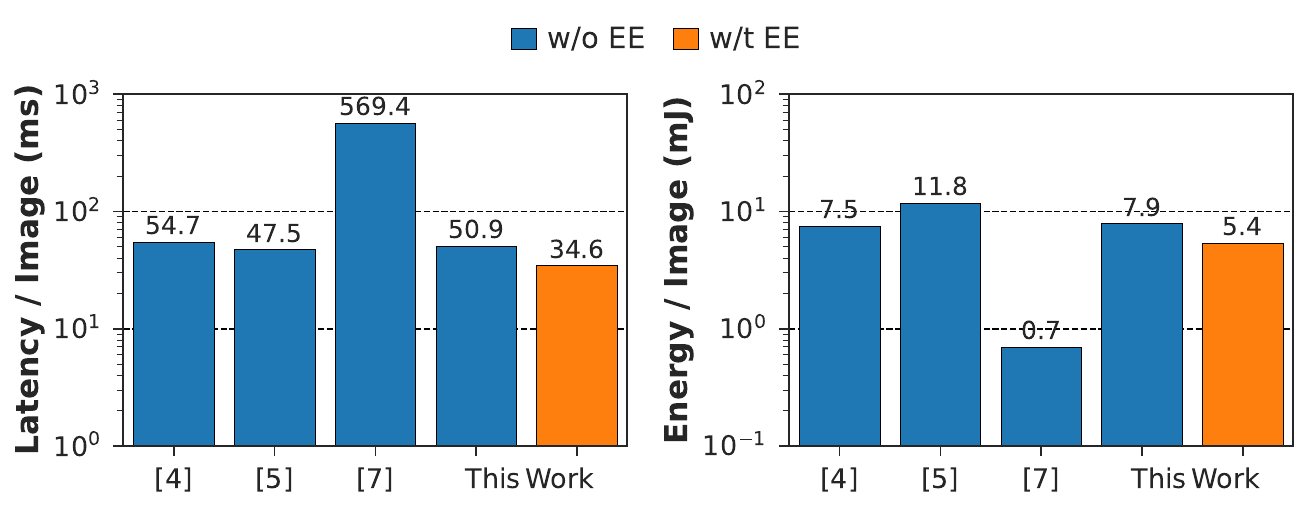} 
    \caption{Average inference latency and energy comparison with and without early exit over the state-of-arts.} 
    \label{fig:perf_inference}
\end{figure}

Fig.~\ref{fig:perf_inference} compares the average inference latency and energy consumption per image of various ODL accelerators, where all results are scaled to $224\times 224$ image resolution for fair comparison. With EE (\(E_s=2\), \(E_c=2\)) enabled, \design's inference latency and energy consumption are reduced roughly by $32\%$. In contrast, prior designs either suffer from long inference latency~\cite{qian20244} or high energy consumption~\cite{wang2022trainer}. Our design achieves the optimal balance between latency and energy efficiency based on early exit mechanism, making it suitable for low-latency and energy-efficient edge inference.

\subsection{Comparison with Existing ODL Chips}

Table~\ref{tab:comparison} compares \design to state-of-the-art ODL chips~\cite{han2021dflnpu,park2021neural,prabhu2022chimera,wang2022trainer,venkataramanaiah202328,qian20244}. \design stands out with its FSL-HDC learning engine, achieving 1.4-2.9 TOPS/W training energy efficiency for ResNet-18. Our chip demonstrates superior FSL training latency and energy efficiency, achieving 6~mJ/image training energy, which outperforms the prior works by $2\times$ to $20.9\times$  with the lowest training latency. This is mainly attributed to the gradient-free and single-pass HDC-based learning paradigm that not only simplifies dataflow, but also significantly reduces the required computational complexity during training.

\begin{figure}[t]
    \centering
    \includegraphics[width=0.78\linewidth]{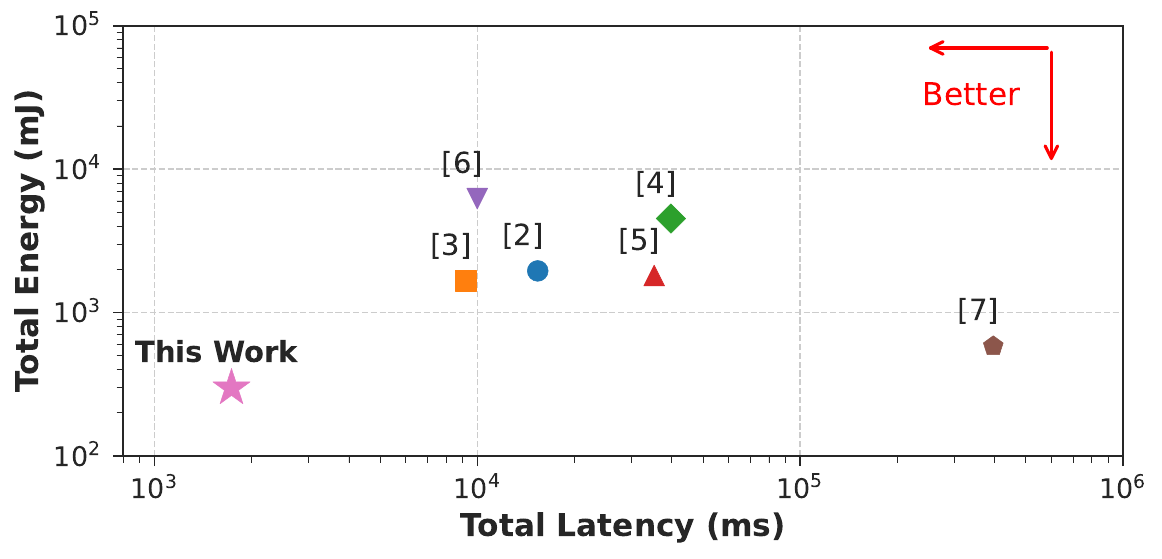} 
    \caption{End-to-end training energy and latency comparison for the 10-way 5-shot FSL tasks. FT baselines use 5 epochs.} 
    \label{fig:e2e_energy_latency}
\end{figure}

We visualize the end-to-end energy and latency  for the 10-way 5-shot FSL tasks in Fig.~\ref{fig:e2e_energy_latency}. \design delivers responsive FSL feedback with just 1.7~sec training time, compared to 9.2 to 396~sec for prior works~\cite{han2021dflnpu,park2021neural,prabhu2022chimera,wang2022trainer,venkataramanaiah202328,qian20244}. At the same time, \design consumes $2\times$–$21\times$ less energy, highlighting its potential for real-time and energy-efficient ODL in edge AI applications.

\section{Conclusion}\label{sec:conclusion}
We propose \design, a 40~nm CMOS accelerator that enables efficient ODL through gradient-free FSL with hyperdimensional computing. Key innovations include weight clustering for parameter-efficient feature extraction, memory-efficient cyclic Random Projection encoding, and single-pass training that eliminates iterative gradient updates. The early exit in feature extraction further improves inference efficiency, while batched training per class boosts training energy and delay efficiency.
\design achieves 6~mJ/image training energy efficiency and $2$–$20.9\times$ speedup over state-of-the-art ODL accelerators while maintaining competitive accuracy, paving the road for energy-efficient edge AI learning systems.

\clearpage

\bibliographystyle{ieeetr}
\bibliography{refs.bib}

@inproceedings{sarangi2021deepscaletool,
  title={DeepScaleTool: A tool for the accurate estimation of technology scaling in the deep-submicron era},
  author={Sarangi, Satyabrata and Baas, Bevan},
  booktitle={International Symposium on Circuits and Systems (ISCAS)},
  pages={1--5},
  year={2021},
}

@inproceedings{he2016deep,
  title={Deep residual learning for image recognition},
  author={He, Kaiming and Zhang, Xiangyu and Ren, Shaoqing and Sun, Jian},
  booktitle={IEEE Conference on Computer Vision and Pattern Recognition},
  pages={770--778},
  year={2016}
}

@article{kanerva2009hyperdimensional,
  title={Hyperdimensional computing: An introduction to computing in distributed representation with high-dimensional random vectors},
  author={Kanerva, Pentti},
  journal={Cognitive Computation},
  volume={1},
  number={2},
  pages={139--159},
  year={2009},
}

@article{thomas2021theoretical,
  title={A theoretical perspective on hyperdimensional computing},
  author={Thomas, Anthony and Dasgupta, Sanjoy and Rosing, Tajana},
  journal={Journal of Artificial Intelligence Research},
  volume={72},
  pages={215--249},
  year={2021}
}

@article{dhar2021odlsurvey,
  title={A survey of on-device machine learning: An algorithms and learning theory perspective},
  author={Dhar, Sauptik and Guo, Junyao and Liu, Jiayi and Tripathi, Samarth and Kurup, Unmesh and Shah, Mohak},
  journal={ACM Transactions on Internet of Things},
  volume={2},
  number={3},
  pages={1--49},
  year={2021},
}

@article{lee2021overview,
  title={An overview of energy-efficient hardware accelerators for on-device deep-neural-network training},
  author={Lee, Jinsu and Yoo, Hoi-Jun},
  journal={IEEE Open Journal of the Solid-State Circuits Society},
  volume={1},
  pages={115--128},
  year={2021},
}

@article{heo2024sp,
  title={SP-PIM: A Super-Pipelined Processing-In-Memory Accelerator With Local Error Prediction for Area/Energy-Efficient On-Device Learning},
  author={Heo, Jaehoon and Kim, Jung-Hoon and Han, Wontak and Kim, Jaeuk and Kim, Joo-Young},
  journal={IEEE Journal of Solid-State Circuits},
  volume={59},
  number={8},
  pages={2671--2683},
  year={2024},
}

@article{cai2020tinytl,
  title={Tinytl: Reduce memory, not parameters for efficient on-device learning},
  author={Cai, Han and Gan, Chuang and Zhu, Ligeng and Han, Song},
  journal={Advances in Neural Information Processing Systems},
  volume={33},
  pages={11285--11297},
  year={2020}
}

@article{heo2022t,
  title={T-PIM: An energy-efficient processing-in-memory accelerator for end-to-end on-device training},
  author={Heo, Jaehoon and Kim, Junsoo and Lim, Sukbin and Han, Wontak and Kim, Joo-Young},
  journal={IEEE Journal of Solid-State Circuits},
  volume={58},
  number={3},
  pages={600--613},
  year={2022},
}

@article{li2021sapiens,
  title={SAPIENS: A 64-kb RRAM-based non-volatile associative memory for one-shot learning and inference at the edge},
  author={Li, Haitong and Chen, Wei-Chen and Levy, Akash and Wang, Ching-Hua and Wang, Hongjie and Chen, Po-Han and Wan, Weier and Khwa, Win-San and Chuang, Harry and Chih, Y.-D. and Chang, Meng-Fan and Wong, H.-S. Philip and Raina, Priyanka},
  journal={IEEE Transactions on Electron Devices},
  volume={68},
  number={12},
  pages={6637--6643},
  year={2021},
}

@inproceedings{khaleghi2022patternet,
  title={Patternet: explore and exploit filter patterns for efficient deep neural networks},
  author={Khaleghi, Behnam and Mallappa, Uday and Yaldiz, Duygu and Yang, Haichao and Shah, Monil and Kang, Jaeyoung and Rosing, Tajana},
  booktitle={ACM/IEEE Design Automation Conference},
  pages={223--228},
  year={2022}
}

@inproceedings{tsai2023scicnn,
  title={SciCNN: A 0-shot-retraining patient-independent epilepsy-tracking SoC},
  author={Tsai, Chne-Wuen and Jiang, Rucheng and Zhang, Lian and Zhang, Miaolin and Wu, Liuhao and Guo, Jiaqi and Yan, Zhongwei and Yoo, Jerald},
  booktitle={IEEE International Solid-State Circuits Conference (ISSCC)},
  pages={488--490},
  year={2023},
}

@article{liu2024high,
  title={A High Accuracy and Ultra-Energy-Efficient Zero-Shot-Retraining Seizure Detection Processor},
  author={Liu, Jiahao and Liu, Xiao and Wang, Xu and Xie, Ziyi and Guo, Chaozheng and Zhong, Zirui and Fan, Jiajing and Qiu, Hui and Xu, Yiming and Qin, Huajing and Long, Yu and Zhou, Yuhong and Shen, Zixuan and Zhou, Liang and Chang, Liang and Liu, Shanshan and Lin, Shuisheng and Wang, Chao and Zhou, Jun},
  journal={IEEE Journal of Solid-State Circuits},
  volume={59},
  number={11},
  pages={3549-3565},
  year={2024},
}

@article{han2021dflnpu,
  title={DF-LNPU: A Pipelined Direct Feedback Alignment-Based Deep Neural Network Learning Processor for Fast Online Learning},
  author={Han, Donghyeon and Lee, Jinsu and Yoo, Hoi-Jun},
  journal={IEEE Journal of Solid-State Circuits},
  volume={56},
  number={5},
  pages={1630--1640},
  year={2021}
}

@article{park2021neural,
  title={A neural network training processor with 8-bit shared exponent bias floating point and multiple-way fused multiply-add trees},
  author={Park, Jeongwoo and Lee, Sunwoo and Jeon, Dongsuk},
  journal={IEEE Journal of Solid-State Circuits},
  volume={57},
  number={3},
  pages={965--977},
  year={2021},
}

@article{prabhu2022chimera,
  title={{CHIMERA: A 0.92-TOPS, 2.2-TOPS/W edge AI accelerator with 2-MByte on-chip foundry resistive RAM for efficient training and inference}},
  author={Prabhu, Kartik and Gural, Albert and Khan, Zainab F. and Radway, Robert M. and Giordano, Massimo and Koul, Kalhan and Doshi, Rohan and Kustin, John W. and Liu, Timothy and Lopes, Gregorio B. and Turbiner, Victor and Khwa, Win-San and Chih, Yu-Der and Chang, Meng-Fan and Lallement, Guénolé and Murmann, Boris and Mitra, Subhasish and Raina, Priyanka},
  journal={IEEE Journal of Solid-State Circuits},
  volume={57},
  number={4},
  pages={1013--1026},
  year={2022},
}

@article{qian20244,
  title={{A 4.69-TOPS/W Training, 2.34-$\mu$J/Image Inference On-Chip Training Accelerator With Inference-Compatible Backpropagation and Design Space Exploration in 28-nm CMOS}},
  author={Qian, Junyi and Ge, Haitao and Lu, Yicheng and Shan, Weiwei},
  journal={IEEE Journal of Solid-State Circuits},
  volume={60},
  number={1},
  pages={298-307},
  year={2025},
}

@article{venkataramanaiah202328,
  title={A 28-nm 8-bit floating-point tensor core-based programmable CNN training processor with dynamic structured sparsity},
  author={Venkataramanaiah, Shreyas Kolala and Meng, Jian and Suh, Han-Sok and Yeo, Injune and Saikia, Jyotishman and Cherupally, Sai Kiran and Zhang, Yichi and Zhang, Zhiru and Seo, Jae-Sun},
  journal={IEEE Journal of Solid-State Circuits},
  volume={58},
  number={7},
  pages={1885--1897},
  year={2023},
}

@article{wang2022trainer,
  title={Trainer: An energy-efficient edge-device training processor supporting dynamic weight pruning},
  author={Wang, Yang and Qin, Yubin and Deng, Dazheng and Wei, Jingchuan and Chen, Tianbao and Lin, Xinhan and Liu, Leibo and Wei, Shaojun and Yin, Shouyi},
  journal={IEEE Journal of Solid-State Circuits},
  volume={57},
  number={10},
  pages={3164--3178},
  year={2022},
}

@article{mao2022experimentally,
  title={Experimentally validated memristive memory augmented neural network with efficient hashing and similarity search},
  author = {Mao, Ruibin and Wen, Bo and Kazemi, Arman and Zhao, Yahui and Laguna, Ann Franchesca and Lin, Rui and Wong, Ngai and Niemier, Michael and Hu, X. Sharon and Sheng, Xia and Graves, Catherine E. and Strachan, John Paul and Li, Can},
  journal={Nature Communications},
  volume={13},
  number={1},
  pages={6284},
  year={2022},
}

@inproceedings{karunaratne2022memory,
  title={In-memory Realization of In-situ Few-shot Continual Learning with a Dynamically Evolving Explicit Memory}, 
  author={Karunaratne, G. and Hersche, M. and Langeneager, J. and Cherubini, G. and Gallo, M. Le and Egger, U. and Brew, K. and Choi, S. and Ok, I. and Silvestre, C. and Li, N. and Saulnier, N. and Chan, V. and Ahsan, I. and Narayanan, V. and Benini, L. and Sebastian, A. and Rahimi, A.},
  booktitle={IEEE European Solid State Circuits Conference (ESSCIRC)},
  pages={105--108},
  year={2022},
}

@inproceedings{ravi2017optimization,
  title={Optimization as a model for few-shot learning},
  author={Ravi, Sachin and Larochelle, Hugo},
  booktitle={International Conference on Learning Representations},
  year={2017}
}

@article{zhou2021device,
  title={On-device learning systems for edge intelligence: A software and hardware synergy perspective},
  author={Zhou, Qihua and Qu, Zhihao and Guo, Song and Luo, Boyuan and Guo, Jingcai and Xu, Zhenda and Akerkar, Rajendra},
  journal={IEEE Internet of Things Journal},
  volume={8},
  number={15},
  pages={11916--11934},
  year={2021},
}

@article{morris2021locality,
  title={Locality-based encoder and model quantization for efficient hyper-dimensional computing},
  author={Morris, Justin and Fernando, Roshan and Hao, Yilun and Imani, Mohsen and Aksanli, Baris and Rosing, Tajana},
  journal={IEEE Transactions on Computer-Aided Design of Integrated Circuits and Systems},
  volume={41},
  number={4},
  pages={897--907},
  year={2021},
  publisher={IEEE}
}

@inproceedings{chowdhury2021few,
  title={Few-shot image classification: Just use a library of pre-trained feature extractors and a simple classifier},
  author={Chowdhury, Arkabandhu and Jiang, Mingchao and Chaudhuri, Swarat and Jermaine, Chris},
  booktitle={Proceedings of the IEEE/CVF International Conference on Computer Vision},
  pages={9445--9454},
  year={2021}
}

@inproceedings{dataset_traffic,
  title={Detection of traffic signs in real-world images: The German Traffic Sign Detection Benchmark},
  author={Houben, Sebastian and Stallkamp, Johannes and Salmen, Jan and Schlipsing, Marc and Igel, Christian},
  booktitle={International Joint Conference on Neural Networks (IJCNN)},
  pages={1--8},
  year={2013},
}

@inproceedings{dataset_flower,
  title={Automated flower classification over a large number of classes},
  author={Nilsback, Maria-Elena and Zisserman, Andrew},
  booktitle={Sixth Indian Conference on Computer Vision, Graphics \& Image Processing},
  pages={722--729},
  year={2008},
}

@article{dataset_cifar100,
  title={Tadam: Task dependent adaptive metric for improved few-shot learning},
  author={Oreshkin, Boris and Rodr{\'\i}guez L{\'o}pez, Pau and Lacoste, Alexandre},
  journal={Advances in Neural Information Processing Systems},
  volume={31},
  year={2018}
}

@inproceedings{oommen2018study,
  title={Study and analysis of various LFSR architectures},
  author={Oommen, Roshni and George, Merin K and Joseph, Sharon},
  booktitle={International Conference on Circuits and Systems in Digital Enterprise Technology (ICCSDET)},
  pages={1--6},
  year={2018},
}

@article{dataset_cifar10,
  title={Learning multiple layers of features from tiny images},
  author={Krizhevsky, Alex and Hinton, Geoffrey},
  year={2009},
  publisher={Toronto, ON, Canada}
}

@inproceedings{deng2009imagenet,
  title={Imagenet: A large-scale hierarchical image database},
  author={Deng, Jia and Dong, Wei and Socher, Richard and Li, Li-Jia and Li, Kai and Fei-Fei, Li},
  booktitle={IEEE Conference on Computer Vision and Pattern Recognition},
  pages={248--255},
  year={2009},
}

@inproceedings{sun2019meta,
  title={Meta-transfer learning for few-shot learning},
  author={Sun, Qianru and Liu, Yaoyao and Chua, Tat-Seng and Schiele, Bernt},
  booktitle={IEEE/CVF Conference on Computer Vision and Pattern Recognition},
  pages={403--412},
  year={2019}
}

@inproceedings{tian2020rethinking,
  title={Rethinking Few-Shot Image Classification: A Good Embedding is All You Need?},
  author={Tian, Yonglong and Wang, Yue and Krishnan, Dilip and Tenenbaum, Joshua B and Isola, Phillip},
  booktitle={European Conference on Computer Vision},
  pages={266--282},
  year={2020}
}

@inproceedings{teerapittayanon2016branchynet,
  title={Branchynet: Fast inference via early exiting from deep neural networks},
  author={Teerapittayanon, Surat and McDanel, Bradley and Kung, Hsiang-Tsung},
  booktitle={International Conference on Pattern Recognition (ICPR)},
  pages={2464--2469},
  year={2016},
}

@article{jo2023locoexnet,
  title={LoCoExNet: Low-cost early exit network for energy efficient CNN accelerator design},
  author={Jo, Joongho and Kim, Geonho and Kim, Seungtae and Park, Jongsun},
  journal={IEEE Transactions on Computer-Aided Design of Integrated Circuits and Systems},
  volume={42},
  number={12},
  pages={4909--4921},
  year={2023},
}

@inproceedings{ilhan2024adaptive,
  title={Adaptive deep neural network inference optimization with eenet},
  author    = {Ilhan, Fatih and Chow, Ka-Ho and Hu, Sihao and Huang, Tiansheng and Tekin, Selim and Wei, Wenqi and Wu, Yanzhao and Lee, Myungjin and Kompella, Ramana and Latapie, Hugo and Liu, Gaowen and Liu, Ling},
  booktitle={IEEE/CVF Winter Conference on Applications of Computer Vision},
  pages={1373--1382},
  year={2024}
}

@article{rahmath2024early,
  title={Early-exit deep neural network-a comprehensive survey},
  author={Rahmath P, Haseena and Srivastava, Vishal and Chaurasia, Kuldeep and Pacheco, Roberto G and Couto, Rodrigo S},
  journal={ACM Computing Surveys},
  volume={57},
  number={3},
  pages={1--37},
  year={2024},
}

@inproceedings{yang2024fsl,
  title={FSL-HDnn: A 5.7 TOPS/W End-to-end Few-shot Learning Classifier Accelerator with Feature Extraction and Hyperdimensional Computing},
  author={Yang, Haichao and Song, Chang Eun and Xu, Weihong and Khaleghi, Behnam and Mallappa, Uday and Shah, Monil and Fan, Keming and Kang, Mingu and Rosing, Tajana},
  booktitle={IEEE European Solid-State Electronics Research Conference (ESSERC)},
  pages={33--36},
  year={2024},
}

@inproceedings{du2018understanding,
  title={Understanding of object detection based on CNN family and YOLO},
  author={Du, Juan},
  booktitle={Journal of Physics: Conference Series},
  volume={1004},
  pages={012029},
  year={2018},
  organization={IOP Publishing}
}

@inproceedings{song2025hybrid,
  title={Hybrid SLC-MLC RRAM Mixed-Signal Processing-in-Memory Architecture for Transformer Acceleration via Gradient Redistribution},
  author={Song, Chang Eun and Bhatnagar, Priyansh and Xia, Zihan and Kim, Nam Sung and Rosing, Tajana S and Kang, Mingu},
  booktitle={Proceedings of the 52nd Annual International Symposium on Computer Architecture},
  pages={1155--1170},
  year={2025}
}

@ARTICLE{11173642,
  author={Song, Chang Eun and Li, Yidong and Ramnani, Amardeep and Agrawal, Pulkit and Agrawal, Purvi and Jang, Sung-Joon and Lee, Sang-Seol and Rosing, Tajana and Kang, Mingu},
  journal={IEEE Journal of Solid-State Circuits}, 
  title={Energy-Efficient Reconfigurable XGBoost Inference Accelerator With Modular Unit Trees via Selective Node Execution and Data Movement}, 
  year={2025},
  volume={},
  number={},
  pages={1-13},
}

\end{document}